%% file: main.tex
\documentclass[conference]{IEEEtran}
\usepackage{cite}

\usepackage{cite}

\ifCLASSINFOpdf
\else
\fi
\usepackage{url}

\usepackage{amsmath}
\usepackage{caption}
\usepackage{subcaption}
\usepackage{multirow}
\usepackage{pgfplots}
\usepgfplotslibrary{statistics}
\usepackage{ulem}
\usepackage{tikz}
\tikzstyle{state}=[
    ellipse,
    thick
]

\usetikzlibrary{shapes}
\definecolor{lavander}{cmyk}{0,0.48,0,0}
\definecolor{violet}{cmyk}{0.79,0.88,0,0}
\definecolor{burntorange}{cmyk}{0,0.52,1,0}
\definecolor{mypurple}{HTML}{bf77f6}
\definecolor{myblue}{HTML}{3776ab}

\begin{document}
%
\title{Time-to-event modeling of subreddits transitions to r/SuicideWatch}


\author{\IEEEauthorblockN{Xueying Liu}
\IEEEauthorblockA{Computational Biology \\
St. Jude Children's Research Hospital\\
Email: Xueying.Liu@stjude.org} \and
\IEEEauthorblockN{Shiaofen Fang}
\IEEEauthorblockA{Computer and Information Science \\
Indiana University - Purdue University\\
Indianapolis\\
Email: sfang@iu.edu}
\and
\IEEEauthorblockN{George Mohler}
\IEEEauthorblockA{Computer Science \\
Boston College\\
Boston\\
Email: mohlerg@bc.edu} \and 
\IEEEauthorblockN{Joan Carlson}
\IEEEauthorblockA{School of Social Work \\
Indiana University - Purdue University\\
Indianapolis\\
Email: joancarl@iu.edu}
\and
\IEEEauthorblockN{Yunyu Xiao}
\IEEEauthorblockA{Population Health Sciences \\
Weill Cornell Medical College\\
Email: yux4008@med.cornell.edu}
}


%


\maketitle

\begin{abstract}
Recent data mining research has focused on the analysis of social media text, content and networks to identify suicide ideation online.  However, there has been limited research on the temporal dynamics of users and suicide ideation.  In this work, we use time-to-event modeling to identify which subreddits have a higher association with users transitioning to posting on r/suicidewatch.  For this purpose we use a Cox proportional hazards model that takes as input text and subreddit network features and outputs a probability distribution for the time until a Reddit user posts on r/suicidewatch.  In our analysis we find a number of statistically significant features that predict earlier transitions to r/suicidewatch.  While some patterns match existing intuition, for example r/depression is positively associated with posting sooner on r/suicidewatch, others were more surprising (for example, the average time between a high risk post on r/Wishlist and a post on r/suicidewatch is 10.2 days).  We then discuss these results as well as directions for future research.

\end{abstract}


%
\IEEEpeerreviewmaketitle

\section{Introduction}
In 2019, approximately 47,500 deaths in the U.S. were attributed to suicide by the Center for Disease Control \cite{cdc}.  Given that suicide can be preventable by early intervention, recent data mining research has focused on the analysis of social media text, content and networks to identify suicide ideation and to better understand social media user risk, trajectories, interactions, and potential interventions \cite{Lindsey2019, Li2021, Xiao2021}.

One line of recent research focuses on detecting suicide ideation in online user content on sites such as Twitter and Reddit \cite{Mbarek19,Odea15,Ji18}.  Other research has focused on modeling data from text messages \cite{Nobles18} and surveys \cite{May20,Xiao19}.  While some studies utilized text based features input into classical machine learning models, more recently deep learning has been used to detect suicide ideation in text data \cite{Ophir20,Tadesse20,Matero19}.  A comprehensive survey on machine learning for suicide detection can be found in \cite{Ji21}, and \cite{Jorge20} provides a survey on mining social networks to improve suicide prevention.

Reddit, in particular, has been the focus of recent data mining research on suicide, as several subreddits such as 'r/suicidewatch' provide forums for individuals thinking about suicide, drug addiction, and/or depression and who may be seeking help from others online.  In \cite{Fraga18}, the authors analyzed discourse patterns of posts and comments on four Reddit online communities including r/depression, r/suicidewatch, r/anxiety and r/bipolar. In \cite{Aladag18},
detection methods were developed for suicide ideation in text on r/suicidewatch and related subreddits and in \cite{Ruch20},
the authors showed how to improve detection on r/suicidewatch by combining graph and language models.  Other work has focused on determining the impact of the COVID-19 pandemic on suicide ideation on Reddit \cite{Low20}, creating an automated question answering system for suicide risk assesment using posts and comments extracted from r/suicidewatch \cite{Alambo19}, and predicting the degree of suicide risk on r/suicidewatch and related subreddits \cite{Zirikly19}.

While a great deal of work has focused on detecting suicide ideation in online posts, there has been limited research on the temporal dynamics of users and suicide ideation.  For example, a user who has suicidal thoughts may post on social media, at which point another may be able to intervene and provide mental health support.  However, it is possible that earlier posts may have contained early indicators that could also have been points for interventions.  In this work our goal is to better understand these earlier events through time-to-event survival analysis of transitions from other subreddit forums to r/suicidewatch.  

In Figure \ref{fig:trajectories}, we show three example post sequences from Reddit that illustrate the type of dynamics we would like to model in the present paper.  The first user posts on r/LongDistance several times, indicating that they feel sad and are having relationship problems due to long distance, the user then posts on r/teenagers a few times expressing their confusion and then later post on r/suicidewatch.  Our goal is to identify which subreddits have a higher association with users transitioning to posting on r/suicidewatch, which text based features are associated with such transitions, and the time between posts from other forums and the first post on r/suicidewatch.  We note that temporal dynamics of suicide ideation on Reddit were considered in \cite{Dutta21}, however the authors analyzed day of week and hour of day trends in the times of posts, rather than analyzing the inter-event time dynamics of transitions to r/suicidewatch.

The outline of the paper is as follows.  In Section \ref{model}, we describe our survival analysis approach, using Cox proportional hazards modeling.  In Section \ref{data}, we provide details on the data we collected from Reddit (r/suicidewatch and connected subreddits).  In Section \ref{results}, we present our results of time-to-event modeling of transitions to r/suicidewatch, including the important features that indicate transitions.  In Section \ref{discussion}, we discuss our results and directions for future work.

\input{trajectories}

\section{Model}
\label{model}

Survival analysis \cite{Klein2006} is a statistical method for analyzing the expected duration until an event occurs. The survival function $S(t)$, defined as $S(t) = P(T \ge t)$,
gives the probability that the time to the event occurs later than an observed time $t$.  The cumulative distribution function (CDF) of the time to event gives the
cumulative probability for a given $t$:
\begin{equation*}
    F(t) = P(T < t) = 1-S(t)
\end{equation*}
The hazard function $h(t)$ is defined as the probability that an event will occur in the time interval $[t, t + \Delta t)$ given that the event has not occurred before:
\begin{equation*}
    h(t) = \lim_{\Delta t\rightarrow 0}\frac{P(t \le T < t + \Delta t | T \ge t)}{\Delta t} = \frac{f(t)}{S(t)},
\end{equation*}
where $f(t)$ is the probability density function (PDF) of the time to event. 

One feature of survival analysis is censoring of event times, as some users observation windows may not be large enough to have fully observed an event outcome.  If for a given user an event of interest has occurred, then the survival time is known (fully observed), whereas for those that the events has not (yet) occurred, we only know that the waiting time exceeds the observation time \cite{Princeton_notes}. These events with unknown survival time are referred to as censored data.  In this study we restrict our analysis to users who post or comment on r/suicidewatch at least once.

The Cox proportional hazards model \cite{Cox1972} is a standard Survival model that allows for the incorporation of covariates.  The idea behind the Cox model is that the log-hazard of an individual is a linear function of a covariate vector \textbf{$x$} and parameter vector \textbf{$\beta$} and a population-level baseline hazard $h_0(t)$ that changes over time.  The Cox's proportional hazard model has the form,
\begin{equation}
    h(t|\mathbf{x}) = h_0(t) \mbox{exp}[g(\mathbf{x})], \quad g(\mathbf{x}) = \mathbf{\beta}^T \mathbf{x},
\label{cox}
\end{equation}
and has been used previously to model transitions to drug addiction and recovery on Reddit \cite{lu2019investigate}.

The Cox model in Equation \ref{cox} is fit to data in two steps\cite{kvamme2019time}.  First, the exponential part is fitted by maximizing the Cox partial likelihood (Equation \ref{Equ:cph}), which does not depend on the baseline hazard, then the baseline hazard $h_0(t)$ is estimated using Breslow's method. 

For observation $i$, let $T_i$ denote the censored event time and $R_i$ denote the set of all observations at risk at time $T_i$. The Cox partial likelihood is defined as
\begin{equation}
    L_{cox} = \prod_i\left(\frac{\mbox{exp}[g(\mathbf{x_i})]}{\sum_{j\in R_i \mbox{exp}[g(\mathbf{x_j})]}}\right)^{D_i},
\label{Equ:cph}
\end{equation}
and the negative partial log-likelihood, which can be used as a loss function, is
\begin{equation}
    \mbox{ll}_{cox} = \sum_i D_i \mbox{log} \left( \sum_{j\in R_i} \mbox{exp} [g(\mathbf{x_j}) - g(\mathbf{x_i})] \right).
    \label{Equ:log_cph}
\end{equation}


Let 
\begin{equation}
    S_x(t) = S(t|X) = P(T>t|X)
    \label{Equ:surv_prob1}
\end{equation} 
be the survival probability at time $t$, then the baseline probability is defined as follows:
\begin{equation}
S_0(t) = e^{-\int_0^{{t}} h_0(t')dt'} = e^{-H_0(t)}.
\label{Equ:base_hazard_prob}
\end{equation} 
For an individual with features $X$, 
\begin{equation}
S_x(t) = e^{-\int_0^t h(t'|x)dt'} = [S_0(t)]^{exp(\beta^T X)}.
\label{Equ:surv_prob2}
\end{equation} 


Let $\hat{\beta}$ be the value of $\beta$ that optimizes (\ref{Equ:cph}) and (\ref{Equ:log_cph}). Then the cumulative baseline hazard function can be estimated by the Breslow estimator\cite{Breslow1972}:

\begin{equation}
    \widehat{H_0(t)} = \sum_{i=1}^n \frac{D_i}{\sum_{j\in R_i \mbox{exp}[g(\mathbf{x_j})]}}.
\end{equation}

Note here $D_i$ is an indicator variable that event $i$ is uncensored, and equals 1 in our model for all $i$.

We use the lifelines CoxPHFitter\footnote{https://lifelines.readthedocs.io/en/latest/fitters/regression/CoxPHFitter.html} in Python to fit the Cox model and estimate coefficients {\bf $\beta$} and baseline hazard.

\section{Data}
\label{data}

The data is collected from Reddit\footnote{https://www.reddit.com}, using PushShift\footnote{https://github.com/pushshift/api} and PRAW\footnote{https://praw.readthedocs.io/en/stable/} APIs. We first obtain a list of users who posted on r/suicidewatch between 1/1/2019 and 12/31/2021. We then randomly sample 2000 users and download their posts over the 3-year period, along with comments from these users that posted on r/suicidewatch and their comments and posts from other subreddits.

After dealing with exceptions on PRAW API and removing deleted posts, we collected more than 163k posts from over 1k users. We retained information including user name, post time, post content, post title, and on which subreddit the post was made.  We then filtered out users with only one post, and posts that occurred after the user already posted on r/suicidewatch. We further cleaned the data by removing special tokens, detecting and translating posts in foreign languages into English, and performing spell check and making corrections. The data we use for analysis throughout this paper contains over 61k posts from 751 users. 


We cut the data at 2020/12/31 23:59:59, and assign posts and comments prior to this time as training data, the rest being the test one. For the users who have posted by the cutoff time but post on r/SuicideWatch afterwards, their corresponding posts in the training data are labeled as \texttt{censored}. There are 129 censored users and 4398 posts.  


\subsection{Suicide ideation detection model}
For each post, we estimated a probability score as to whether a post contained language associated with suicide ideation using a pre-trained model \cite{Kaggle}. The model is trained on text data collected from r/suicidewatch and r/depression and utilizes a LSTM neural network based on text embeddings with `suicidal' vs. `non-suicidal' binary labels.  We use this model to estimate a score between 0 and 1 that represents the probability that a post in our dataset is associated with suicide.  We then define posts to be `high risk' if the score is higher than 0.95, and low otherwise.

\subsection{Summary statistics and figures}
In Table \ref{tab:subreddit_score}, we provide average suicidal scores of some most frequently posted subreddits, where on r/suicidewatch, the score is noticeably higher.  In Table ~\ref{tab:summary1-1} we provide a summary of the number of posts of each user, where the average number of posts is 53.4 per user.  In Table ~\ref{tab:summary1-2} we provide a summary of the length of posts with high and low suicidal scores, where we find that high risk scores are associated with longer posts.  We also find that posts with high suicidal scores are more likely to occur on weekends (Friday and Saturday).
Figure \ref{fig:hist1} shows a histogram of users' posts, where $10\%$ of users have only 2 posts and $76\%$ of users have less than 50 posts each.  Figure \ref{fig:hist2} shows the distribution of inter-event times between posts on other sub-reddits and r/suicidewatch.  More than 3000 of the posts were made within 3.5 days of posting on r/suicidewatch. The longest waiting time before posint on r/suicidewatch is 698 days.  
\input{subreddit_score}

\begin{table}[h]
\begin{subtable}{.45\linewidth}
    \centering
    \begin{tabular}{|l|r|}
    \hline
    Max & 838 \\
    \hline
    Min & 2\\
    \hline
    Mean & 81.34 \\
    \hline
    \end{tabular}
    \caption{Number of posts and comments per user.}
    \label{tab:summary1-1}
    \end{subtable}
\begin{subtable}{.5\linewidth}
\centering
\begin{tabular}{|l | c | c| }
\hline 
& High & Low \\
\hline
Mean length & 539.18 & 164.38\\
\hline
$\%$ on weekend & 27.69 & 27.94 \\ 
\hline
\end{tabular}
\caption{Posts and comments grouped by suicidal scores.}
\label{tab:summary1-2}
\end{subtable}
\caption{Summary of data 1}
\label{tab:summary1}
\end{table}


\input{no_post_hist}


\input{time_hist}

\subsection{Topic models}
\label{subsec-tm}
We analyze the topic models of text data by first utilizing SentenceTransformers \cite{reimers-2019-sentence-bert} - a BERT-based pretrained model that derives semantically meaningful sentence embeddings. We then perform KMeans with the embeddings that associate with high suicidal score (i.e. $>$ 0.95). We search for the optimal $K$ using the ``elbow" method, which suggests $K=4$. Moreover, we extract keywords of each topic with the help of spaCy library in Python. The keywords are displayed in Table \ref{tab:keywords_topic}. 

\begin{figure}[t]
    \centering
    \includegraphics[scale=0.4]{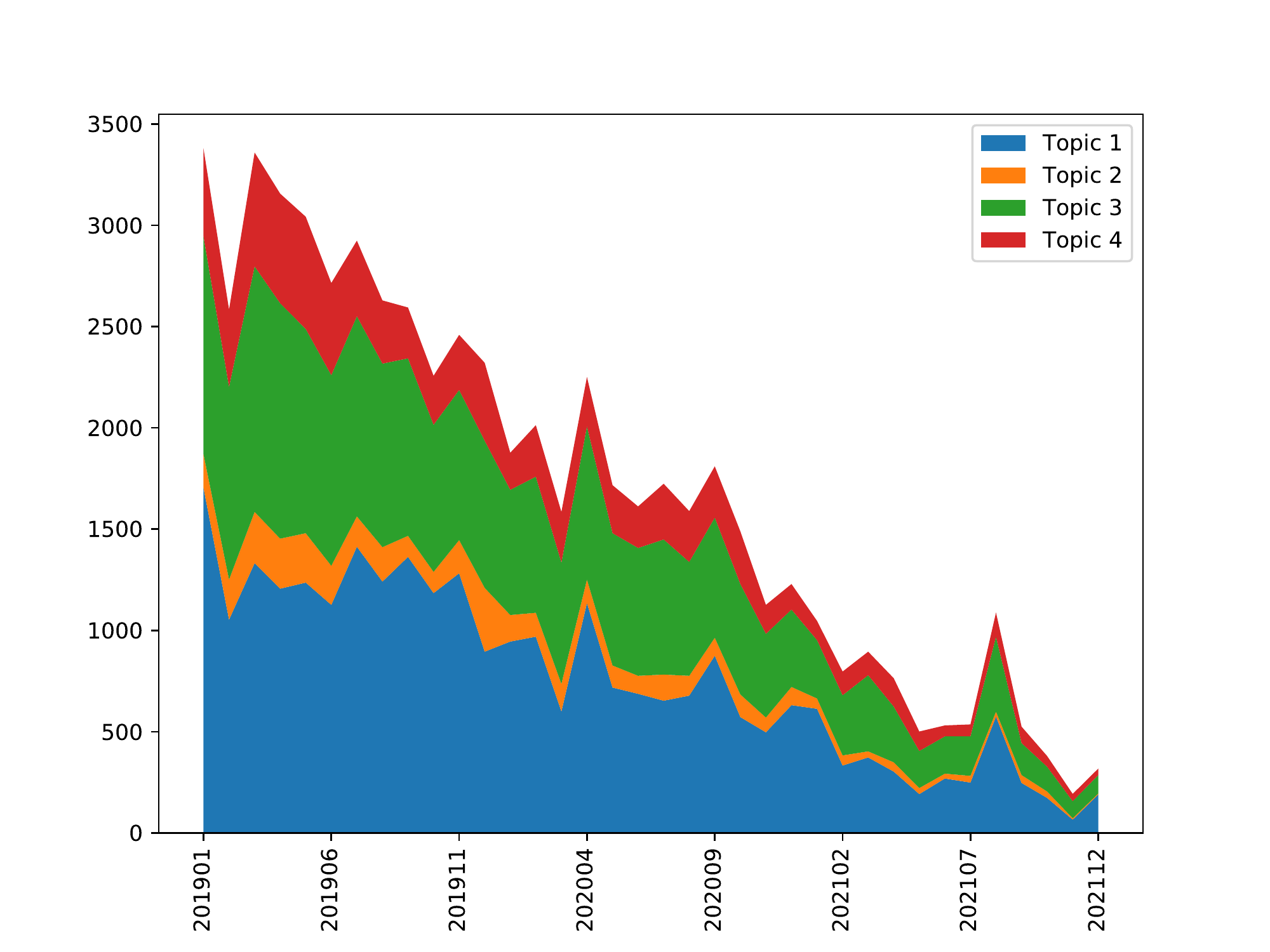}
    \caption{Popularity of each topic over time.}
    \label{fig:stream_graph}
\end{figure}

\input{topic_keyword}
Figure \ref{fig:stream_graph} demonstrates the popularity of each topic over the 3-year observed timeframe. 


\subsection{Feature Selection}
\subsubsection{Keyword expansion}
We use keyword expansion to determine a list of keywords related to suicide.  Starting with a manually selected list of 39 keywords, we then use cosine similarity of word vectors to find the 10 most similar words to each.  We use the word2vec implementation in gensim \cite{rehurek_lrec} to create the 100-dimensional word vector representations.  This process results in a keyword list of length 331.  We next create dummy variables that indicate if a post contains each of these keywords. Figure \ref{fig:hist4} shows the histogram of number of keywords present in a post or comment. 40 most frequently occurred keywords are displayed in Table \ref{tab:top_keyword}.

\input{keyword_list}

\input{keyword_hist}

\subsubsection{Sources connecting to subreddit r/suicidewatch}
We select the top 50 frequent subreddits (excluding r/suicidewatch) and create dummy variables that indicate if a post is from one of these subreddits. In Figure \ref{fig:hist3}, we show the most frequently posted 15 subreddits. On average, the data contains 13.75 posts and comments from each subreddit.

\input{subreddit_hist}


\subsubsection{Index of topics}
The topic indices obtained from part \ref{subsec-tm} are transferred into dummy variables. 
\section{Results}
\label{results}

  We fit a Cox proportional hazards model to our data, with a penalizer term of 5, and within the summary table of results, we focus on the variables with a p-value less than 0.05. In Table \ref{tab:coef}, we show the coefficients of these statistically significant variables.  The subreddits with the highest coefficients (indicating sooner transition to r/suicidewatch) include r/selfharm, r/Wishlist, r/awakened, r/BreakUps and r/MadeOfStyrofoam (which is a forum for selfharm discussion).  Subreddits associated with longer time-to-event intervals between an initial post and a subsequent post on r/suicidewatch include r/LivestreamFail, r/AvPD, r/ftm and r/PurplePillDebate (a forum to discuss sex and gender issues).  While a majority of the keywords were not statistically significant, both `pain' and `life' were associated with shorter time-to-event periods, as was the high risk category based on the suicide detection model described above. The keyword `women' appeared to be associated with a longer time-to-event interval. Topic feature is not statistically significant.

\begin{figure}[h]
    \centering
    \includegraphics[scale =0.45]{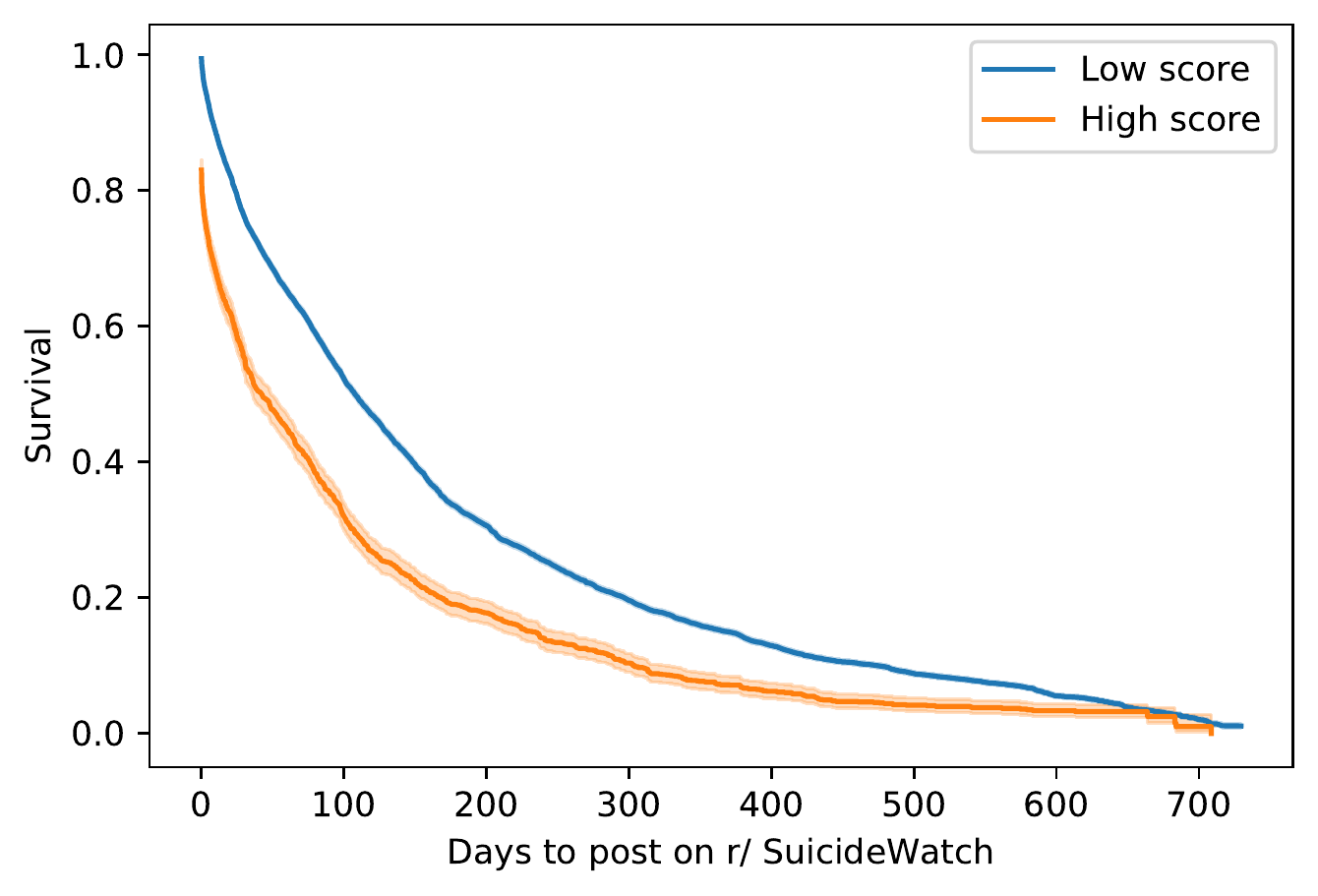}
    \caption{Kalpan Meier estimates by suicidal score}
    \label{fig:KM}
\end{figure}

    In Figure \ref{fig:KM}, we display the estimated Kaplan Meier curve for the distribution of time-to-event disaggregated by high vs. low suicidal scores.  Here we find that the time-to-event distribution has a shorter tail for higher risk scores, indicating that posts with high risk scores are associated with subsequent r/suicidewatch posts occuring sooner.  In Figure \ref{fig:transition1}, we display the transition network from other subreddits (yellow indicating a positive association, blue indicating a negative association) to r/suicidewatch along with the average transition time between the final post preceding a post on r/suicidewatch and their first post on r/suicidewatch. On each edge, the number represents the average number of days between subreddit and r/suicidewatch posts when the suicidal score is high (low). No post is found from r/cats, r/MortalKomat, r/sweden and r/Eminem with a high suicidal score.

    
    We predict expected remaining lifetime of each censored user and compute the concordance between prediction and ground truth, the model obtains a concordance index of 0.5123. We also compute AUC by dividing the entire future into 30-day interval. We label an interval $1$ if transition to r/SuicideWatch occurred in the interval. This gives us an AUC of 0.8214. 
    

    
    

\begin{table}[h]
    \centering
    \begin{tabular}{|l|r|l|r|}
    \cline{1-2}
    \multicolumn{2}{|l|}{\textbf{Subreddit indicators}} & \multicolumn{2}{}{}\\
    \hline
        depression & 0.06& LivestramFail &  -0.28  \\ 
        teenagers & 0.05  & fireemblem & -0.08 \\
        relationship$\_$advice & 0.05 &  MortalKombat& -0.12\\
        awakened & 0.15 & AvPD & -0.37\\
        selfharm & 0.09 & Sweden & -0.23\\
         MadeOfStyrofoam & 0.11 & Traaa...nnnns & -0.1\\
          Wishlist & 0.17 & ftm & -0.19 \\
          
        BPD & 0.08 &  PurplePillDebate & -0.29\\
                \cline{3-4}

          Eminem & 0.14 &  \multicolumn{2}{c}{}\\
          cats & 0.1 & \multicolumn{2}{c}{} \\
         unpopularopinion & 0.06 &\multicolumn{2}{c}{}\\
        BreakUps & 0.12 & \multicolumn{2}{c}{}\\
    \hline
    \multicolumn{2}{|l|}{\textbf{Keyword indicators}} & \multicolumn{2}{|l|}{\textbf{Suicidal score}} \\
    \hline
        ``pain" & 0.04 & score & 0.04 \\\cline{3-4}
        ``women" & -0.07 & \multicolumn{2}{c}{} \\ 
        ``life" & 0.02 & \multicolumn{2}{c}{} \\
    \cline{1-2}
    
    \end{tabular}
    \caption{Coefficients of significant variables.}
    \label{tab:coef}
\end{table}

\input{transition_fig}

\section{Discussion}
\label{discussion}

In this research, we collected a large corpus of suicide related posts from r/suicidewatch, along with earlier posts made by users on other subreddits.  We then fit a Cox proportional hazards model to predict the time-to-event between earlier posts and later posts on r/suicidewatch.  We found statistically significant features using indicators for subreddit, keyword, or suicide risk.  While some patterns match existing intuition, for example r/teenagers and r/relationship$\_$advice are positively associated with posting sooner on r/suicidewatch, others were more surprising.  For example, the average time between a high risk post on r/Wishlist and a consecutively following post on r/suicidewatch is 10.2 days (less than the 97.2 day average time between events on r/depression and r/suicidewatch).  Our results indicate potential points of earlier intervention and analysis of associated subreddits to suicide may yield new hypotheses for suicide researchers to investigate.

Future research may improve upon our work in several ways.  While we used a Cox proportional hazards model, a deep learning based survival model \cite{zhao2021bertsurv} may yield improvements to accuracy.  Also, we did not explore the social networks of individuals and how interactions on the network may be predictive of transitions to suicide ideation. In future work we hope to analyze posting behavior of network connections and people whom a given person interacts with, and determine how those interactions may act to protect against or lead to higher risk of suicide ideation observed online.


 \section*{Acknowledgments}

 This research was supported in part by AFOSR MURI grant FA9550-22-1-0380 and a seed grant from the IUPUI Institute of Integrative AI.



%

\bibliographystyle{ieeetr}
\bibliography{bibliography}

\end{document}

%% file: trajectories.tex
\begin{figure}
\centering
\resizebox{0.4\textwidth}{!}{%

\begin{tikzpicture}
\draw[->] (-1,-3.)--(7.5,-3.) node[below right] {$t$};
\node(user1) at (-0.8,-3.5) [draw = purple, thick, rectangle, text width=1cm]
{User 1 }; 
\node[text width=0.3cm, thick] at (1.7,-3.3) 
   { \textcolor{purple}{$t_1$}};
\node[text width=0.3cm, thick] at (4.5,-3.3) 
   { \textcolor{purple}{$t_2$}};
\node[text width=0.3cm, thick] at (6.85,-3.3) 
   { \textcolor{purple}{$t_3$}};

\node(time1) at (1.7,-3.8) [draw = purple, rectangle, text width = 2.1cm]
{r/LongDistance};
\node(time2) at (4.6,-3.8) [draw = purple, rectangle, text width = 1.5cm]
{r/teenagers};
\node(time3) at (6.85,-3.8) [draw = purple, rectangle, text width = 2.1cm]
{r/SuicideWatch};


\node(text1) at (0.9,-0.5) [draw = purple,rectangle callout,text width=2cm] 
  {feel wonderful to have him around a lot};
\draw[dashed] (1.7,-3.)--(1.7,-1.75) ;

\node(text2) at (4,-1.) [draw = purple,rectangle callout,text width=3.cm] 
  {I'm about to be 20 in four months. It feels weird.};
\draw[dashed] (4.6,-3.)--(4.6,-2.25) ;

\node(text3) at (6.,1.1) [draw = purple,rectangle callout,text width=3.5cm] 
  {It's been a few years since I've been at the edge, but I feel it again... I'm having a hard time.};
\draw[dashed] (6.85,-3.)--(6.85,-0.5) ;


\draw[->] (-1,-9.)--(7.5,-9.) node[below right] {$t$};
\node(user2) at (-0.8,-9.5) [draw = orange, thick, rectangle, text width=1cm]
{User 2}; 

\node(text1) at (0.3,-5.5) [draw = orange,rectangle callout,text width=3.5cm] 
  {... We got through every thing and by the time it was over I had fallen in love with this series all over again....};
\draw[dashed] (1.15,-9.)--(1.15,-7) ;

\node(text2) at (3.6,-7.6) [draw = orange,rectangle callout,text width=4cm] 
  {... Everything just feels awful and miserable. I'm driving everyone away with my awful symptoms. ...};
\draw[dashed] (4.4,-9.)--(4.4,-8.8) ;

\node(text3) at (6.,-5.5) [draw = orange,rectangle callout,text width=4.5cm] 
  {...Not only was he perfect, but he only stopped loving me cause of my Borderline Personality Disorder meltdown and rages I'd have.};
\draw[dashed] (6.9,-9)--(6.9,-7.1) ;

\node[text width=0.3cm, thick] at (1.3,-9.3) 
   { \textcolor{orange}{$t_1$}};
\node[text width=0.3cm, thick] at (4.4,-9.3) 
   { \textcolor{orange}{$t_2$}};
\node[text width=0.3cm, thick] at (6.9,-9.3) 
   { \textcolor{orange}{$t_3$}};
   
\node(time1) at (1.15,-9.8) [draw = orange, rectangle, text width = 2.2cm]
{r/stevenuniverse};
\node(time2) at (4.4,-9.8) [draw = orange, rectangle, text width = 0.8cm]
{r/BPD};
\node(time3) at (7,-9.8) [draw = orange, rectangle, text width = 2.1cm]
{r/SuicideWatch};



\draw[->] (-1,-15)--(7.5,-15) node[below right] {$t$};
\node(user3) at (-0.8,-15.5) [draw = blue, thick, rectangle, text width=1cm]
{User 3}; 

\node[text width=0.3cm, thick] at (1,-15.3) 
   { \textcolor{blue}{$t_1$}};
\node[text width=0.3cm, thick] at (3.6,-15.3) 
   { \textcolor{blue}{$t_2$}};
\node[text width=0.3cm, thick] at (6.65,-15.3) 
   { \textcolor{blue}{$t_3$}};

\node(time1) at (1.,-15.85) [draw = blue, rectangle, text width = 1.2cm]
{r/hockey};
\node(time2) at (3.6,-15.85) [draw = blue, rectangle, text width = 1.8cm]
{r/AskReddit};
\node(time3) at (6.65,-15.85) [draw = blue, rectangle, text width = 2.1cm]
{r/SuicideWatch};

\node(text1) at (0.,-12.1) [draw = blue,rectangle callout,text width=3.2cm] 
  {I remember dealing with so much racism when I played,... Seeing stuff like this gives me hope and makes me smile};
\draw[dashed] (1.,-15)--(1.,-14.1) ;

\node(text2) at (2.85,-13.5) [draw = blue,rectangle callout,text width=1.75cm] 
  {I always feel like I was asking for to much};
\draw[dashed] (3.6,-15)--(3.6,-14.9) ;

\node(text3) at (6.,-12.25) [draw = blue,rectangle callout,text width=3.9cm] 
  {I threw up blood and started experiencing awful chest pain, I haven’t told anyone. I’m tired of feeling alone,... scared, ... hiding from everyone. I feel so judged.};
\draw[dashed] (7.1,-15)--(7.1,-14.4) ;


\end{tikzpicture}
}
\caption{Posting sequences of 3 Reddit users.}
\label{fig:trajectories}
\end{figure}

%% file: subreddit_score.tex
\begin{table}[h]
    \centering
    \begin{tabular}{|l|r| l | r|}
    \hline
    Subreddit & Average score & Subreddit & Average score \\
    \hline
      AskReddit   & 0.1730 & dankmemes & 0.1631 \\
      teenagers & 0.1455 & AmItheAsshole & 0.2027\\
       {\bf SuicideWatch} &{\bf 0.8036} & trees & 0.0958 \\
       memes & 0.1485 & FortNiteBR & 0.0963\\
       depression  & 0.6200 & selfharm & 0.4478\\
       AskOuija & 0.1416 & awakened & 0.4240 \\
        relationship$\_$advice & 0.3657 & Advice & 0.4374\\
        unpopularopinion & 0.1425 & NoFap & 0.2781\\
 \hline
    \end{tabular}
    \caption{Average suicidal score of posts on popular subreddits.}
    \label{tab:subreddit_score}
\end{table}

%% file: no_post_hist.tex
\input{datasets}
\begin{figure}
\centering
\resizebox{0.7\columnwidth}{!}{%

\begin{tikzpicture}
    \begin{axis}[
            ybar,
            bar width=.1cm,
            xticklabel style = {font = \tiny},
            ymin=0,ymax=105,
            yticklabel style = {font = \tiny},
            xlabel = {\small Number of posts and comments per user},
            xtick pos=bottom,ytick pos=left
        ]
        \addplot [fill = myblue!60, draw = myblue!60, 
        hist = {
            bins = 300, 
            data min = 0,
        }]
        table[x expr = \coordindex, y = {Count}] {hist1.csv};
    \end{axis}

\end{tikzpicture}
}
\caption{Histogram of users' posts and comments.}
\label{fig:hist1}
\end{figure}

%% file: datasets.tex
\pgfplotstableread[row sep=\\,    col sep=&]{
    Post ct & Keyword ct  \\
   38265 & 0. \\
   13605 & 0.84 \\
   4711 & 1.68 \\
   1910 & 2.52 \\
   968 & 3.36 \\
   599 & 4.2 \\
   0 & 5.04 \\
   362 & 5.88 \\
   239 & 6.72 \\
   123 & 7.56 \\
   82 & 8.4 \\
   63 & 9.24 \\
   0 & 10.08 \\
   36 & 10.92 \\
   35 & 11.76 \\
   24 & 12.6 \\
   13 & 13.44 \\
   11 & 14.28 \\
   0 & 15.12 \\
   12 & 15.96 \\
   8 & 16.8 \\
   5 & 17.64 \\
   2 & 18.48 \\
   3 & 19.32 \\
   0 & 20.16 \\
   2 & 21 \\
   2 & 21.84\\
   1 & 22.68 \\
   1 & 23.52 \\
   0 & 24.36 \\
   1 & 25.2 \\
   0 & 26.04 \\
   1 & 26.88 \\
   1 & 27.72 \\
   0 & 28.56 \\
   0 & 29.4 \\
   0 & 30.24 \\
   0 & 31.08 \\
   0 & 31.92 \\
   0 & 32.76\\
   0 & 33.6 \\
   0 & 34.44 \\
   0 & 35.28 \\
   0 & 36.12 \\
   0 & 36.96 \\
   0 & 37.8 \\
   0 & 38.64 \\
   1 & 39.48 \\
   0 & 40.32 \\
   0 & 41.16 \\
   2 & 42 \\
    }\mykeyword

\pgfplotstableread[row sep=\\,    col sep=&]{
    subreddit & count  \\
    AskReddit     & 5309    \\
    teenagers     & 2433   \\
    memes   & 1330  \\
    depression   & 687  \\
    AskOuija & 638 \\
    relationship\_advice & 562\\
        unpopularopinion  & 527  \\
    dankmemes & 457\\
        AmItheAsshole & 428 \\
            trees & 389\\
    FortNiteBR & 370 \\
        selfharm & 361 \\
    awakened & 329\\
    Advice & 321 \\
    NoFap & 315 \\
    }\mysubreddit

  \pgfplotstableread[row sep = \\,     col sep = &]{
  feel & 2134 \\
  work & 1084 \\
  god & 448 \\
  pain & 385 \\
  mental & 363 \\
  depression & 342 \\
  anxiety & 279 \\
  health & 256 \\
  death & 235 \\
  therapy & 203 \\
  therapist & 173 \\
  cry & 157 \\
  drugs & 123 \\
  positive & 122 \\
  emotional & 120 \\
  lonely & 110 \\
  guilty & 91 \\
  drug & 90 \\
  illness & 89 \\
  weed & 87 \\
  alcohol & 82 \\
  building & 81 \\
  mood & 76 \\
  career & 75 \\
  psychiatrist & 74\\
  smoke & 73 \\
  drunk & 68 \\
  mg & 68 \\
  pressure & 67 \\
  dose & 65 \\
  jump & 63 \\
  toxic & 63 \\
  guilt & 63 \\
  homeless & 61 \\
  smoking & 61 \\
  addiction & 59 \\
  religion & 57  \\
  terrified & 53 \\
  exercise & 52 \\
  mixed & 46\\
  }\mykeywordlist

  \pgfplotstableread[row sep = \\,     col sep = &]{
  Weekday & High & Low \\
  Sun & 13.3936 & 14.1627 \\
  Mon & 15.8997 & 14.2548  \\
  Tues & 15.5711 & 15.6221 \\
  Wed & 14.2563 & 14.1116 \\
  Thurs & 14.1331 & 14.5122 \\
  Fri & 12.4486 & 13.5610 \\
  Sat & 14.2974 & 13.7757 \\
  }\myweekday

%% file: time_hist.tex
\input{datasets}
\begin{figure}[h]
\centering
\resizebox{0.75\columnwidth}{!}{%

\begin{tikzpicture}
    \begin{axis}[
            ybar,
            bar width=.1cm,
            xticklabel style = {font = \tiny},
            ymin=0,ymax=6000,
            ytick={0,500, 1000,...,5500,6000},
            yticklabel style = {font = \tiny},
            xlabel = {\small Time (days) to event},
            xtick pos=bottom,ytick pos=left
        ]
        \addplot [fill = myblue!60, draw = myblue!60]table[x=Days,y=Count, col sep=comma]{hist2.csv};
    \end{axis}

\end{tikzpicture}
}
\caption{Histogram of time to event.}
\label{fig:hist2}
\end{figure}

%% file: topic_keyword.tex
\begin{table}[h]
    \centering
    \begin{tabular}{|l|r|}
    \hline
    Topic & Keywords \\
    \hline
    Topic 1 & right, tear, know\\
    \hline
    \multirow{2}{*}{Topic 2}&think, sending, affected, died, unjustified, death,\\
    &help, threaten, pills, life, traumatising, emotionally,...\\
    \hline
    Topic 3 & tried, turning, fix, way, find\\
    \hline
    \multirow{2}{*}{Topic 4} & longer, like, want, future, realized, world, care, depression, \\
    &time,method, meant, planned, fought, struggled, \\
    &torture, hope,...\\
       \hline
    \end{tabular}
    \caption{Keywords extracted from posts with high suicidal score.}
    \label{tab:keywords_topic}
\end{table}

%% file: keyword_list.tex
\begin{table}[h]
    \centering
    \begin{tabular}{|l|r|l|r|l|r|}
    \hline 
    word & count & word & count & word & count\\
    \hline
like & 8823 & soon & 565 &  abuse & 329\\
  good & 3462 & pain & 550 & depressed & 326\\
  way & 2708  & relationship & 534 & therapy & 306  \\
  life & 2220 & damn & 527 & red & 251 \\
 thanks & 1899 & check & 457  &  cry & 238 \\
  work & 1657 & anxiety & 455 & therapist & 228 \\
  friends & 1254 & health & 448  &  account & 224 \\
   friend & 985 & kid & 423 & upset & 198 \\
   idea & 799 & important & 381 &  weed & 186\\
   place & 778 & death & 372 & ending & 182 \\
   god & 696  & eat & 356 & toxic & 174 \\
   mental & 690  & type & 354 & emotional & 169 \\ 
  women & 589 & bring & 336 &  party & 164\\
  using & 569 & & & & \\
  \hline
    \end{tabular}
    \caption{Top 40 most frequent keywords.}
    \label{tab:top_keyword}
\end{table}

%% file: keyword_hist.tex
\input{datasets}
\begin{figure}
\centering
\resizebox{0.75\columnwidth}{!}{%

\begin{tikzpicture}
    \begin{axis}[
            ybar,
            bar width=.2cm,
            width=.5\textwidth,
            height=.4\textwidth,
            xtick={0,5,10,15,20,25,30,35,40,45},
            xticklabel style = {font = \tiny},
            ymin=0,ymax=40000,
            yticklabel style = {font = \tiny},
            ylabel = {\small Number of posts},
            xtick pos=bottom,ytick pos=left
        ]
        \addplot [fill = myblue!60, draw = myblue!60]table[x=Keyword ct,y=Post ct]{\mykeyword};
    \end{axis}
    \node [below=1.cm, right = 1.75cm] at (0,0) {\small Number of keywords in a post};

\end{tikzpicture}
}
\caption{Histogram of number of keywords.}
\label{fig:hist4}
\end{figure}

%% file: subreddit_hist.tex
 \input{datasets}
\begin{figure}[h]
\centering
\resizebox{0.75\columnwidth}{!}{%

\begin{tikzpicture}
    \begin{axis}[
            ybar,
            bar width=.3cm,
            width=.5\textwidth,
            height=.4\textwidth,
            symbolic x coords={AskReddit, teenagers, memes, depression,  
             AskOuija, relationship\_advice, unpopularopinion, dankmemes, 
            AmItheAsshole,trees, FortNiteBR, selfharm,awakened, Advice, NoFap},
            xtick=data,
            xticklabel style = {rotate=-90, text width = 1.5cm, font = \tiny},
            ymin=0,ymax=5500,
            ytick={0,500, 1000,...,5000,5500},
            yticklabel style = {font = \tiny},
            ylabel = {\small Number of posts},
            xtick pos=bottom,ytick pos=left
        ]
        \addplot [fill = myblue!60, draw = myblue!60]table[x=subreddit,y=count]{\mysubreddit};
    \end{axis}
    \node [below=1.72cm, right = 2.5cm] at (0,-.5) {\small Subreddits};

\end{tikzpicture}
}
\caption{15 mostly posted subreddits.}
\label{fig:hist3}
\end{figure}

%% file: transition_fig.tex
\begin{figure}
\centering
        \resizebox{0.8\columnwidth}{!}{%

\begin{tikzpicture}


\node [state, minimum size = 1.5cm, fill = red!60, text = white] (center) at (4,-4){\large \begin{tabular}{c}
     Suicide\\Watch
\end{tabular}};
 
\node[state, minimum size = 1.25 cm, fill = yellow, text = blue, rotate = 25] (A) at (.5,-6.2){depression};
 
\node[state, minimum size = 1.25 cm, fill = yellow, text = blue] (B) at (4.5,1){teenagers};
 
\node[state, minimum size = 1.25 cm, fill = yellow, text = blue, rotate = -45] (C) at (7.8,-8.5){relationship$\_$advice};

\node[state, minimum size = 1.25 cm, fill = yellow, text = blue, rotate = -15] (D) at (2,-1.8){Cats};

\node[state, minimum size = 1.25 cm, fill = yellow, text = blue, rotate = -30] (E) at (8.,-7){selfharm};

\node[state, minimum size = 1.25 cm, fill = blue!80, text = yellow] (F) at (7,0.5){fireemblem};

\node[state, minimum size = 1.25 cm, fill = yellow, text = blue] (G) at (-.5,-3){awakened};

\node[state, minimum size = 1. cm, fill = yellow, text = blue] (H) at (8.5,-4){BPD};

\node[state, minimum size = 1.25 cm, fill = yellow, text = blue] (I) at (0.5,-4.4){Wishlist};

\node[state, minimum size = 1.25 cm, fill = yellow, text = blue, rotate = 60] (J) at (2,-7.5){\begin{tabular}{c} unpopular \\ opinions \end{tabular}};

\node[state, minimum size = 1.25 cm, fill = blue!80, text = yellow, rotate = 33] (K) at (6.8,-1.8){MortalKombat};

\node[state, minimum size = 1.25 cm, fill = blue!80, text = yellow] (L) at (2,-0.5){ftm};

\node[state, minimum size = 1.25 cm, fill = blue!80, text = yellow, rotate = -60] (M) at (5.2,-7.5){AvPD};

\node[state, minimum size = 1.25 cm, fill = blue!80, text = yellow] (N) at (3,3){Sweden};

\node[state, minimum size = 1.25 cm, fill = blue!80, text = yellow, rotate = -45] (O) at (1.,2.){LivestreamFail};

\node[state, minimum size = 1.25 cm, fill = yellow, text = blue, rotate = 50] (P) at (6.5,3){Eminem};

\node[state, minimum size = 1.25 cm, fill = blue!80, text = yellow, rotate = 15] (Q) at (8.5,-2.5){\begin{tabular}{c} Traaaaaaa \\ nnnnnnnnnns \end{tabular}};

\node[state, minimum size = 1.25 cm, fill = blue!80, text = yellow, rotate = -90] (R) at (4,-8.5){PurplePillDebate};

\node[state, minimum size = 1.25 cm, fill = yellow, text = blue, rotate = -15] (S) at (8.5,-5.5){MadeOfStyrofoam};

\node[state, minimum size = 1.25 cm, fill = yellow, text = blue, rotate = -20] (T) at (0,-1.5){BreakUps};

\draw (A) to[left] node[above, rotate=35]{\tiny 97.2(84)}(center) ;
 
\draw (B) to[bend right =5] node[above, near start, rotate = 85]{\tiny 85.3(112.6)}(center);
 
\draw (C) to[left]node[above , rotate=-45]{\tiny 68.5(98.1)} (center);
 
\draw (D) to[bend left=10] node[below, rotate = -45]{\tiny (111.5)}(center) ;
 
\draw (E) to[left] node[above, near end, rotate = -30]{\tiny 68.1(53.6)}(center);
 
\draw (F) to[bend right=5]node[above, near start, rotate = 60]{\tiny 288.9(279.7)} (center);

\draw (G) to[bend left=10] node[below, rotate = -15]{\tiny 10.7(23.7)}(center) ;
 
\draw (H) to[right] node[above]{\tiny 81.1(79.8)}(center);
 
\draw (I) to[left]node[above]{\tiny 10.2(115.35)}(center);

\draw (J) to[left] node[above, rotate=60]{\tiny 196.9(176)}(center) ;
 
\draw (K) to[left] node[above, rotate = 40]{\tiny (241.)}(center);
 
\draw (L) to[bend left=10]node[above,near start, rotate = -45]{\tiny 95.7(100.6)} (center);

\draw (M) to[bend left=5] node[above, rotate=-60]{\tiny 447.8(291.7)}(center) ;
 
\draw (N) to[left] node[below, near start, rotate = -80]{\tiny (158.9)}(center);
 
\draw (O) to[bend left=15]node[above, near start, rotate = -55]{\tiny 574.9(169.9)} (center);

\draw (P) to[right] node[above, rotate = 75]{\tiny (26.8)}(center) ;
 
\draw (Q) to[left] node[above, rotate = 15]{\tiny 247.5(165)}(center);
 
\draw (R) to[left]node[below, rotate=-90]{\tiny 681.1(639.8)} (center);

\draw (S) to[left]node[above,  rotate = -15]{\tiny 131(77.1)} (center);

\draw (T) to[right]node[below,  rotate = -20]{\tiny 117.1(60.9)} (center);





\end{tikzpicture}
}
\caption{Average days from the most recent subreddit transitions to r/ SuicideWatch with a high (low) suicidal score}
\label{fig:transition1}
\end{figure}
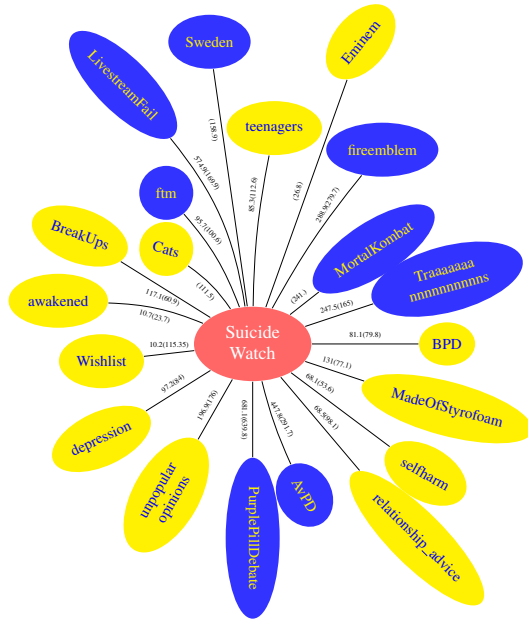